\documentstyle[12pt]{article}

\newcommand{\beq}{\begin{equation}}
\newcommand{\eeq}{\end{equation}}

\begin{document}

\title{Efficient fault-tolerant quantum computing}
\author{Andrew M. Steane \\
{\small Department of Atomic and Laser Physics, University of Oxford,}\\
{\small Clarendon Laboratory, Parks Road, Oxford OX1 3PU, England.}}
\date{October 1998}
\maketitle

\begin{abstract}
Fault tolerant quantum computing methods which work with efficient quantum
error correcting codes are discussed. Several new techniques are introduced
to restrict accumulation of errors before or during the recovery. Classes of
eligible quantum codes are obtained, and good candidates exhibited. This
permits a new analysis of the permissible error rates and minimum overheads
for robust quantum computing. It is found that, under the standard noise
model of ubiquitous stochastic, uncorrelated errors, a quantum computer need
be only an order of magnitude larger than the logical machine contained
within it in order to be reliable. For example, a scale-up by a factor of
22, with gate error rate of order $10^{-5}$, is sufficient to permit large
quantum algorithms such as factorization of thousand-digit numbers.
\end{abstract}


{\bf keywords} Quantum error correction, quantum computing, fault tolerant
\newpage

The future of computing lies in fault-tolerant architectures. This is true
both of classical computing methods, and of future quantum computers. In
both cases the reason is that it is much easier to build a device with
significant imperfections, but with the flexibility to work around them,
than it is to build an essentially ``perfect'' physical device (one whose
chances of failure during any required task are acceptably small). There is
a profound dichotomy at work here, between the power of information
processing, and the effects of random noise and imprecision. The central
point is that information processing itself provides powerful techniques to
protect against information loss.

It is a striking feature of biology that from the molecular level (e.g.
transcription of DNA), up to the operations of organs or the whole organism,
the operating principle involves imperfect structures with built-in
self-correction, rather than close to perfect structures. In classical
computing methods, self-correction has played a part, but it has not so far
been such a central and all-pervasive ingredient. However, this appears set
to change, since as circuit elements get smaller it becomes at once harder
to make them precisely and easier to make sufficiently many that a fraction
can be devoted to error-correction at little cost \cite{teramac}.

In quantum computing \cite{RPP} the need for error correction is paramount
right from the start, since it appears that it may be {\em impossible} to
find a physical system which could be sufficiently precise and isolated to
constitute a useful `bare' quantum computer. Here, by a `bare' quantum
device we mean one whose physical operation only involves elements (qubits,
gates, etc.) essential to the logical structure of the task in hand, and by
a `useful' quantum computer we mean a general-purpose quantum computational
device which could tackle computing tasks not readily solvable by other
means (such as classical computing). It seemed up until only a few years ago
that this difficulty ruled out useful quantum computers altogether, since it
was unknown how to achieve error correction in quantum processing. We now
know how to realise quantum error correction \cite
{Shor1,Steane1,Cald1,Steane2,Steane3,Knill,Benn,concat,intro} and fault
tolerant quantum circuits \cite
{ft,Kita,DiVS,Steane97,KLZ96,AhBO96,resil,Pres,req,Gottft}, at a cost in the
size and speed of the computer. Thus, useful quantum computing appears to be
allowed by the laws of nature, and there are two questions which present
themselves:

\begin{enumerate}
\item  {\em What is the maximum quantum computing power achievable in a
system of given dimension and noise rates?}

\item  {\em How is the maximum attained?}
\end{enumerate}

These questions are important both from the point of view of our
understanding of fundamental physics, and from the practical point of view
of building quantum computers. They are the subject of this paper. Their
practical importance is large because quantum computers are so hard to make:
controllable qubits are a precious resource which we wish to use as
efficiently as possible. Up till now, proposed fault tolerant quantum
computing methods have been inefficient because they are based on
inefficient quantum coding, in which only a single qubit is stored in each
block of $n$ qubits \cite{KLZ96,resil,Pres,req}. This results in a physical
quantum computer which is a hundred to a thousand times larger than logical
machine embedded within it, if we wish to run a large quantum algorithm such
as factorization of hundred- or thousand-digit numbers \cite{Pres,req}.

It is known that more efficient codes exist \cite
{Steane1,Cald1,Steane2,Steane3,Gott,CRSS,CaldGF4,eCSS}. Knill \cite{Knilleb}
discussed ways to find operations on general codes and recently Gottesman 
\cite{Gottft} derived a complete set of fault tolerant operations which can
work with efficient quantum codes. However, these methods require further
refinement if we are to profit by them, otherwise the greater complexity of
the operations lowers the error tolerance, thus offsetting the gain in
coding efficiency.

This paper discusses fault tolerant computing using efficient quantum codes,
including specific example codes and estimates of the noise rates which can
be tolerated. Section \ref{s:1} considers universal sets of quantum gates,
and sections \ref{s:2},\ref{s:3} discuss a universal set of fault tolerant
operations for Calderbank Shor Steane (CSS) quantum codes satisfying certain
requirements. Section \ref{s:4} obtains classes of codes satisfying the
requirements, and section \ref{s:5} discusses fault tolerant recovery for
these codes. The analysis of the whole method yields an estimate for the
error rate which can be tolerated and the total computer size needed. The
main conclusion is remarkable: to run a given large quantum algorithm, with
given tolerated error rates in the memory and elementary operations, the
physical quantum computer can be an order of magnitude smaller than
previously thought. This represents a significant step forward both for the
practical prospects of quantum computation, and towards understanding the
fundamental physics underlying questions (1) and (2) enumerated above.

\section{Universal gates}

\label{s:1}

The following notation will be adopted. The single-qubit operators $X$ and $Z
$ are the Pauli operators $\sigma_x$, $\sigma_z$, respectively, and $Y = XZ$%
. We use $H$ for the single-qubit Hadamard operation, $R = HZ$ for the
rotation through $\pi/2$ about the $y$ axis of the Bloch sphere, and $P$ for
the rotation through $\pi/2$ about the $z$ axis (phase shift of $\left| {1}
\right>$ by $i$).Thus $R^2 = Y$, $P^2=Z$ and $(HPH)^2 = X$. The general
phase shift of $\left| {1} \right>$ by $\exp(i\phi)$ will be written $P(\phi)
$, so $P=P(\pi/2)$, $Z=P(\pi)$, etc.

A controlled $U$ operation is written $^C\! U$, so for example $^C\! X$ is
controlled-not, and $T \equiv ^{CC}\!\! X$ is the Toffoli gate.

For operations on bare qubits, the most commonly considered universal set of
quantum gates is $\{ U(\theta, \phi), ^C\! X \}$ where $U(\theta, \phi)$ is
a general rotation of a single qubit. However, this is not a useful set to
consider for the purpose of finding fault-tolerant gates on encoded qubits,
because $U(\theta, \phi)$ is not readily amenable to fault-tolerant methods.

Five slightly different proposals for fault-tolerant universal sets have
been put forward. All involve the {\em normalizer group}, generated by $\{X,
Z, H, P,\, ^C\! X\}$ \cite{resil,Gottft}. Since $Z=P^2$ and $X=HZH$ it is
sufficient to have $\{H, P,\, ^C\! X\}$ to cover this group. The normalizer
group is not sufficient for universal quantum computation, however, nor even
for useful quantum computation, since it can be shown that a quantum
computer using only operations from this group can be efficiently simulated
on a classical computer \cite{pKnill}. To complete the set a further
operation must be added.

\begin{enumerate}
\item  Shor \cite{ft} proposed adding the Toffoli gate, making the universal
set $\{H,P,^{C}\!X,T\}$ (or $\{R,P,^{C}\!X,T\}$ which is equivalent since $%
R=HP^{2}$). Obviously, $^{C}\!X$ can be obtained from $T$, but this does not
reduce the set since Shor's method to obtain $T$ assumes that $^{C}\!X$ is
already available.

\item  Knill, Laflamme and Zurek \cite{KLZ96} proposed $\{P,^{C}\!P,^{C}\!X\}
$ together with the ability to prepare the encoded (or `logical') states $%
\left| {+}\right\rangle _{L}\equiv \left( \left| {0}\right\rangle
_{L}+\left| {1}\right\rangle _{L}\right) /\sqrt{2}$, $\left| {-}%
\right\rangle _{L}\equiv \left( \left| {0}\right\rangle _{L}-\left| {1}%
\right\rangle _{L}\right) /\sqrt{2}$. This can be shown to be sufficient
since preparation of $\left| {\pm }\right\rangle _{L}$ together with $P$ and 
$X$ can produce $H$, and $^{C}\!P$ and $^{C}\!X$ suffice to produce\footnote{%
Note the rule of thumb that a controlled rotation through $\theta$ about
some axis can be obtained by combining $^C\! X$ and single-bit rotations
through $\theta/2$; a controlled-controlled rotation through $\theta$ can be
obtained by combining $^C\! X$ and controlled $\theta/2$ rotations \cite
{gates}.} $^{CC}\!\!Z$, which with $H$ makes $^{CC}\!\!X=T$.

\item  The same paper also considers $\{H,P,^{C}\!X,^{C}\!P\}$.

\item  The same authors subsequently \cite{resil} proposed $\{H,P,^{C}\!X\}$
combined with preparation of $\left| {\pi /8}\right\rangle _{L}=\cos (\pi
/8)\left| {0}\right\rangle _{L}+\sin (\pi /8)\left| {1}\right\rangle _{L}$.
The latter is prepared by making use of the fact that it is an eigenstate of 
$H$, and once prepared is used to obtain a $^{C}\!H$ operation, from which
the Toffoli gate can be obtained.

\item  Gottesman \cite{Gottft} showed that $^{C}\!X$, combined with the
ability to measure $X,Y$ and $Z$, is sufficient to produce any operation in
the normalizer group. The universal set is completed by $T$, following Shor.
\end{enumerate}

Of the above methods, (1) is a useful starting point and will be used
extensively in what follows, but we will need to generalize it to $[[n,k,d]]$
codes storing more than one qubit per block. (2) will be considered also,
though the codes for which it works turn out to be non-optimal. (3) will not
be adopted because the implementation of $^{C}\!P$ involves repeated
recoveries against $X$ errors before a single recovery against $Z$ errors is
possible. This means that $Z$ errors accumulate for a long time before they
can be corrected, and the resulting error tolerance is low. (4) will not be
adopted because it is slow, requiring 12 preparations of $\left| {\pi /8}%
\right\rangle _{L}$ for every Toffoli gate, and the preparation is itself
non-trivial. (5) is important because measurement of $X$, $Y$ and $Z$ can be
performed fault-tolerantly for any stabilizer code, not just $[[n,1,d]]$
codes. Gottesman also proposed the use of measurements and whole-block
operations to swap logical qubits between and within blocks. Thus the
Gottesman methods rely heavily on measurement, which might be thought to be
disadvantageous. However, all the methods (1) to (5) involve measurement
and/or state preparation to implement the Toffoli gate $T$. Since any useful
quantum computation must make significant use of $T$ (otherwise it could be
efficiently simulated classically), methods (1), (2) and (5) are all roughly
equivalent in this regard. For example, the speed of Shor's algorithm to
factorize integers is limited by the Toffoli gates required to evaluate
modular exponentials \cite{Beck,Pres}.

So far we have some universal sets of gates, but we lack a construction to
show how to achieve the particular operations we might need in a given
quantum algorithm. Typically quantum algorithms are built up from the
normalizer group and the Toffoli gate, combined with rotations of single
qubits. Preskill \cite{Pres} provides a construction using two Toffoli
gates, measurements and a $P$ gate to obtain $P(\phi )$ where $\cos \phi =3/5
$. By repeated use of this and the $\pi /2$ rotations it is easy to build
any other rotation to the requisite precision. Note that this trick
generalizes as follows: if the $P$ gate is replaced by $P(\alpha )$, then
the overall result is $P(\phi )$ where $\cos \phi =(6+10\cos \alpha
)/(10+6\cos \alpha )$.

\section{Fault-tolerant operations for CSS codes}

\label{s:2}

In the list described in the previous section, (1) to (4) gave
fault-tolerant operations for certain $[[n,1,d]]$ Calderbank Shor Steane
(CSS) quantum codes \cite{Steane1,Cald1,Steane2,Pres}; this section will
discuss the generalization to $[[n,k,d]]$ codes. (5) gave fault-tolerant
operations for any stabilizer code\cite{Gott,CRSS,CaldGF4,eCSS}; this
section will give details on the application to CSS codes, and section \ref
{s:3} will introduce further refinements.

The CSS quantum codes are those whose stabilizer generators separate into $X$
and $Z$ parts \cite{Gott,eCSS}. We restrict attention to these codes, rather
than any stabilizer code, because they permit a larger set of
easy-to-implement fault tolerant operations, and their coding rate $k/n$ can
be close to that of the best stabilizer codes. The CSS codes have the
property that the zeroth quantum codeword can be written as an equal
superposition of the words of a linear classical code $C_0$, 
\begin{equation}
\left| {0} \right>_L = \sum_{x \in C_0} \left| {x} \right>,  \label{cw0}
\end{equation}
where $\left| {x} \right>$ is a product state, $x$ is a binary word, and the
other codewords are formed from cosets of $C_0$. Let $\tilde{D}$ be the $k
\times n$ binary matrix of coset leaders, then the complete set of encoded
basis states is given by 
\begin{equation}
\left| {u} \right>_L = \sum_{x \in C_0} \left| {x + u \cdot \tilde{D}}
\right>,  \label{uL}
\end{equation}
where $u$ is a $k$-bit binary word. We will adopt the convention throughout
that symbols with a tilde, such as $\tilde{D}$, refer to binary matrices.
This will avoid confusion between the Hadamard operator $H$ and a parity
check matrix $\tilde{H}$.

We define an operation to be `legitimate' if it maps the encoded Hilbert
space onto itself. We define an operation to be `fault tolerant' if it does
not cause errors in one physical qubit to propagate to two or more qubits in
any one block. Bitwise application of a two-bit operator is defined to mean
the operator is applied once to each pair of corresponding bits in two
blocks, and similarly for bitwise three-bit operations across three blocks.
Legitimate bitwise operations are fault tolerant.

The following notation will be useful. The bar as in $\overline{U}$ is used
to denote the operation $U$ occurring in the encoded, i.e. logical, Hilbert
space, thus $_{L}\!\left\langle {u}\right| \overline{U}\left| {v}%
\right\rangle _{L}=\left\langle {u}\right| U\left| {v}\right\rangle $. A
block of $n$ physical qubits stores $k$ logical qubits. The notation $M_{i}$%
, where $i$ is an $n$-bit binary word, means a tensor product of
single-qubit $M$ operators acting on those physical qubits identified by the
1s in $i$ (for example $X_{101}=X\otimes I\otimes X$). The notation $%
\overline{M}_{u}$, where $u$ is a $k$-bit binary word, means a tensor
product of $\overline{M}$ operators acting on the logical qubits identified
by the 1s in $u$.

Consider a CSS code as defined in eq. (\ref{uL}). Then the encoded $X$ and $Z
$ operators are given by 
\begin{eqnarray}
\overline{X}_u &=& X_{u \cdot \tilde{D}}  \label{Xbar} \\
\overline{Z}_u &=& Z_{u \cdot (\tilde{D}\tilde{D}^T)^{-1} \cdot \tilde{D}}
\label{Zbar}
\end{eqnarray}
Equation (\ref{Xbar}) follows immediately from the code construction, eq. (%
\ref{uL}), and (\ref{Zbar}) follows from $Z_i = \overline{Z}_{i \cdot \tilde{%
D}^T}$, which can be seen from the observation that $Z_i$ changes the sign
of $\left| {u} \right>_L$ whenever $u \cdot \tilde{D}$ fails the parity
check $i$.

We will now examine operators which cannot be expressed as products of $%
\overline{X}$ and $\overline{Z}$.

{\bf Lemma 1.} For $[[n,1,d]]$ codes where all words in $\left| {0} \right>_L
$ have weight $r_0 \mbox{ mod } w$, and all words in $\left| {1} \right>_L$
have weight $r_1 \mbox{ mod } w$, bitwise application of the following are
legitimate: $P(2\pi/w)$, $^C\! P(4\pi/w)$, $^{CC}\!\! P(8\pi/w)$, and
achieve respectively $\overline{P}(2r\pi/w)$, $^C\! \overline{P}(4r\pi/w)$, $%
^{CC}\!\! \overline{P}(8r\pi/w)$, where $r=r_1 - r_0$.

{\em Proof:} for clarity we will take $r_{0}=0$ and $r_{1}=r$, the proof is
easily extended to general $r_{0}$. The argument for $^{C}\!P(4\pi /w)$ was
given in \cite{KLZ96}, but we shall need it for $^{CC}\!\!P(8\pi /w)$, so we
repeat it here. Consider $^{C}\!P(4\pi /w)$ applied to a tensor product of
two codewords. Let $x,y$ be binary words appearing in the expressions for
the two codewords, and let $a$ be the overlap (number of positions sharing a
1) between $x$ and $y$. Let $|x|$ denote the weight of a word $x$. Then $%
2a=|x|+|y|-|x+y|$. There are three cases to consider. First if $x,y\in C_{0}$
then $|x|=0\mbox{ mod }w,\;|y|=0\mbox{ mod }w$ and $|x+y|=0\mbox{ mod }w$ so 
$2a=0\mbox{ mod }w$ from which $a=0\mbox{ mod }w/2$. Therefore the
multiplying factor introduced by the bitwise operation is $1$. If $x\in C_{0}
$ and $y\in C_{1}$ then $x+y\in C_{1}$ so $|x|=0\mbox{ mod }w,\;|y|=|x+y|=r%
\mbox{ mod }w$ so $2a=0\mbox{ mod }w$ again. If $x,y\in C_{1}$ then $x+y\in
C_{0}$ so $a=r\mbox{ mod }w/2$ and the multiplying factor is $\exp (ir4\pi
/w)$. The resulting operation in the logical Hilbert space is therefore $%
^{C}\!\overline{P}(4r\pi /w)$.

Next consider $^{CC}\!\! P(8\pi/w)$ applied to a tensor product of three
codewords. Let $x,y,z$ be words appearing in the three codeword expressions,
and $a,b,c$ be the overlap between $x$ and $y$, $y$ and $z$, and $z$ and $x$%
, respectively. Let $d$ be the common overlap of $x,y$ and $z$, so 
\begin{equation}
|x+y+z| = |x| + |y| + |z| - 2a - 2b - 2c + 4d.
\end{equation}
There are four cases to consider. If $x,y,z \in C_0$ then $d=0 \mbox{ mod }
w/4$. If $x,y \in C_0, z \in C_1$ then $|x+y+z| = |z|$, $2a=2b=2c=0 \mbox{
mod } w$ from the argument just given, therefore $d=0 \mbox{ mod } w/4$. If $%
x \in C_0$, $y,z \in C_1$ then $x+y+z \in C_0$, $2a=2c=0 \mbox{ mod } w$
while $2b = 2r \mbox{ mod } w = |y| + |z|$ so again $d=0 \mbox{ mod } w/4$.
If $x,y,z \in C_1$ then $x+y+z \in C_1$, $2a=2b=2c=2r \mbox{ mod } w$,
therefore $d=r \mbox{ mod } w/4$. The overall effect is that of the
operation $^{CC}\!\! \overline{P}(8r\pi/w)$. \hfill $\Box$

Lemma 1 applied to codes with $w=8$ or more provides a quicker way to
generate the Toffoli gate than previously noted, and also provides an extra
single-bit gate $\overline{P}(7\pi /8)$. The latter can be used to generate
further rotations using the generalized two-Toffoli method referred to at
the end of section \ref{s:1}. The Lemma 1 concept generalizes to $%
^{ccc}\!P(16\pi /w)$ and so on, but the codes for which this is useful (i.e.
having $w\ge 16$) are either inefficient or too unwieldy to produce good
error thresholds.

Gottesman \cite{Gottft} provides an elegant way to find some fault-tolerant
operations. Let ${\cal G}$ be the group generated by $n$-bit products of $I,
X, Y, Z$ where $I$ is the identity, let $M$ be a member of ${\cal G}$, and
let $S$ be the stabilizer. Then for operations $U$ satisfying $U M
U^{\dagger} \in G$ $\forall M \in S$, $U$ is legitimate as long as $U M
U^{\dagger} \in S$. This formalism permits one to prove the following
straightforwardly:

{\bf Lemma 2.} Bitwise $^C\! X$ is legitimate for all CSS codes.

{\bf Lemma 3.} Bitwise $H$ and $^C\! Z$ are legitimate for any $%
[[n,2k_c-n,d]]$ CSS code obtained from a $[n,k_c,d]$ classical code which
contains its dual.

{\bf Lemma 4.} Let $C$ be a $[n,k_c,d]$ classical code which contains its
dual, and for which the weights of the rows of the parity check matrix are
all integer multiples of 4. Then bitwise $P$ is legitimate for the $%
[[n,2k_c-n,d]]$ CSS code obtained from $C$.

An alternative proof of these lemmas will emerge as we examine the effect of
the relevant operations.

Bitwise $^C\! X$ acts as follows: 
\begin{eqnarray}
^C\! X_{{\rm bitwise}} \left| {u} \right>_L \left| {v} \right>_L &=& \sum_{x
\in C_0} \sum_{y \in C_0} \left| {x+u\cdot \tilde{D}} \right> \left| {%
y+v\cdot \tilde{D} + x + u \cdot\tilde{D}} \right> \\
&=& \sum_{x \in C_0} \sum_{y \in C_0} \left| {x+u\cdot \tilde{D}} \right>
\left| {y+(u+v)\cdot \tilde{D}} \right> \\
&=& \left| {u} \right>_L \left| {u+v} \right>_L.  \label{CXbar}
\end{eqnarray}
This is $^C\! \overline{X}$ from each logical qubit in the first block to
the corresponding one in the second.

Bitwise $^C\! H$ acts as follows on $\left| {u} \right>_L$: 
\begin{equation}
H_{{\rm bitwise}} \sum_{x \in C_0} \left| {x + u \cdot \tilde{D}} \right> =
\sum_{y \in C_0^{\perp}} (-1)^{u \tilde{D} y^T} \left| {y} \right>.
\end{equation}
If $C_0^{\perp}$ contains its dual $C_0$, as required for lemma 3, then $%
\tilde{D}$ and $C_0$ together generate $C_0^{\perp}$, so this can be written 
\begin{equation}
H_{{\rm bitwise}} \left| {u} \right>_L = \sum_{v=0}^{2^k-1} \sum_{x \in C_0}
(-1)^{u \tilde{D} \tilde{D}^T v^T} \left| {x + v \cdot \tilde{D}} \right> =
\sum_{v=0}^{2^k-1} (-1)^{u \tilde{D} \tilde{D}^T v^T} \left| {v} \right>_L
\label{Hbar}
\end{equation}
where to simplify the power of $(-1)$ we used the fact that $C_0$ is
generated by the parity check matrix $C_0^{\perp}$, so $u \tilde{D}$
satisfies the parity check $x \in C_0$. Equation (\ref{Hbar}) is a Hadamard
transform acting in the logical Hilbert space when $\tilde{D} \tilde{D}^T = 
\tilde{I}$, and is a closely related transformation when $\tilde{D} \tilde{D}%
^T \ne \tilde{I}$.

Using the above ideas it is easy to show that bitwise $^C\! Z$ produces, for
codes satisfying lemma 3, 
\begin{equation}
^C\! Z_{{\rm bitwise}} \left| {u} \right>_L \left| {v} \right>_L = (-1)^{u 
\tilde{D} \tilde{D}^T v^T} \left| {u} \right>_L \left| {v} \right>_L.
\label{CZbar}
\end{equation}

We will prove lemma 4 by showing that all the quantum codewords have $|x + u
\cdot \tilde{D}| = |u \cdot \tilde{D}| \mbox{ mod } 4$, so the weights
modulo 4 of the components in (\ref{uL}) depend on $u$ but not on $x$. The
effect of bitwise $P$ will therefore be to multiply $\left| {u} \right>_L$
by the phase factor $i^{|u \cdot \tilde{D}|}$.

The zeroth codeword is composed from the code $C_0 = C^{\perp}$ generated by 
$\tilde{H}$, the parity check matrix of $C$. Let $y$ and $z$ be two rows of $%
\tilde{H}$, then the conditions of the lemma guarantee $|y| = 0 \mbox{ mod }
4$ and $|z| = 0 \mbox{ mod } 4$. Furthermore, since $C$ contains its dual,
each row of $\tilde{H}$ satisfies all the checks in $\tilde{H}$, so $y$ and $%
z$ have even overlap $2m$. Therefore $|y+z| = 4m \mbox{ mod } 4 = 0 \mbox{
mod } 4$, therefore $|x| = 0 \mbox{ mod } 4$ for all words in $\left| {0}
\right>_L$. Next consider a coset, formed by displacing $C_0$ by the vector $%
w=u \cdot \tilde{D}$. Since this coset is in $C$ it also satisfies all the
checks in $\tilde{H}$, therefore its members have even overlap with any $x
\in C_0$. Hence if $|w| = r \mbox{ mod } 4$ then $|x + w| = r \mbox{ mod } 4$
for all the terms in the coset, which proves the lemma.

The case $\tilde{D}\tilde{D}^{T}=\tilde{I}$, which leads to a simple effect
for bitwise $H$, also simplifies bitwise $P$. If $\tilde{D}\tilde{D}^{T}=%
\tilde{I}$ then every row of $\tilde{D}$ has odd overlap with itself (i.e.
odd weight) and even overlap with all the other rows. Using an argument
along similar lines to the one just given, we deduce that the effect is the $%
\overline{P}^{r}$ operator applied to every logical qubit in the block,
where $r$ is the weight the relevant row of $\tilde{D}$.

\section{Measurements and the Toffoli gate}

\label{s:3}

Our set of fault tolerant operations now contains sufficient to generate the
group ${\cal G}$ of encoded $\overline{I}, \overline{X}, \overline{Y}, 
\overline{Z}$ operations on individual logical qubits, and the normalizer
group on whole blocks ($k$ logical qubits) at a time, for the lemma 4 codes.
It remains to extend the normalizer group to individual encoded qubits, and
to find a fault tolerant Toffoli gate. For the former, we adopt Gottesman's 
\cite{Gottft} proposal of switching logical qubits into otherwise empty
blocks, applying whole-block operations, then switching back. For the
latter, we use inter-block switching together with Shor's \cite{ft}
implementation of the Toffoli gate, as simplified by Preskill \cite{Pres}.
That the Shor technique works for lemma 4 (and lemma 3) codes follows from
the following:

{\bf Lemma 5.} For CSS codes in which bitwise $^C\! Z$ is legitimate,
bitwise $^{CC}\!\! Z$ is legitimate when operating on two control blocks in
the logical Hilbert space, and a target block in the space spanned by $n$%
-bit `cat' states $\left| {000 \cdots 0} \right> \pm \left| {111 \cdots 1}
\right>$. If bitwise $^C\! Z$ has the effect $\left| {u} \right>_L \left| {v}
\right>_L \rightarrow (-1)^{uv^T} \left| {u} \right>_L \left| {v} \right>_L$%
, then bitwise $^{CC}\!\! Z$ has the effect $\left| {u} \right>_L \left| {v}
\right>_L \left| {a} \right> \rightarrow (-1)^{a(uv^T)} \left| {u} \right>_L
\left| {v} \right>_L \left| {a} \right>$, where $a = 0$ or $1$ and $\left| {a%
} \right>$ means the $n$-bit state $\left| {000 \cdots 0} \right>$ or $%
\left| {111 \cdots 1} \right>$ accordingly.

{\em Proof:} Consider eq. (\ref{CZbar}) and expand $\left| {u} \right>_L
\left| {v} \right>_L$ into a sum of $2n$-bit product states $\left| {x}
\right>\left| {y} \right>$. The bitwise $^C\! Z$ operator can only have the
effect (\ref{CZbar}) if the overlap of $x$ and $y$ is the same, modulo 2,
for every term in the sum. Therefore the bitwise $^{CC}\!\! Z$ operator as
described in lemma 5 produces the same number of $Z$ operations on the cat
state, modulo 2, for every term in the corresponding expansion, and the
effect is as described. \hfill $\Box$

Gottesman's switching and swapping techniques make much use of the ability
to measure $\overline{X}$ or $\overline{Z}$ operators fault tolerantly. A
method to perform such measurements was deduced by DiVincenzo and Shor \cite
{DiVS}, based on preparation of verified `cat' states $\left| {000\cdots 0}
\right> + \left| {111\cdots 1} \right>$. However, the preparation and
verification of these states involves many elementary gates, and the
measurement must be repeated to ensure reliability. These operations take a
considerable number of elementary gates and time steps, during which errors
accumulate. This significantly reduces the tolerance on error rates in the
computer. Our next ingredient is an important trick to circumvent this
problem:

{\bf Lemma 6.} For any stabilizer code, measurement of any operator $%
\overline{X}_u$ or $\overline{Y}_u$ or $\overline{Z}_u$ can be performed at
no cost by merging it with the recovery operation.

\begin{figure}[tbp]
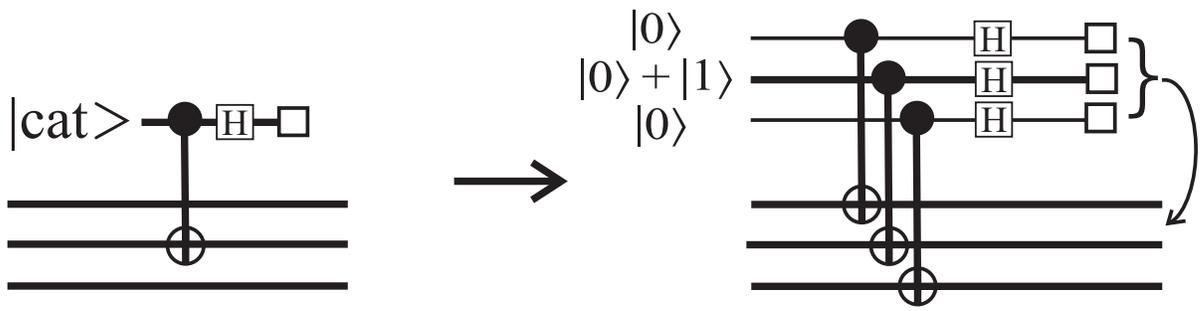

\caption{Method to measure $\overline{X}_{010}$ while simultaneously
extracting the syndrome for $Z$ errors. The left hand network is that
proposed in \protect\cite{DiVS}, it is replaced by the right hand network.
The horizontal lines represent logical qubits, the boxes represent the
measurement of the ancilla, from which a syndrome and the required $%
\overline{X}_{010}$ measurement can be deduced. The final arrow represents
correction of the deduced $Z$ errors. The necessary repetition discussed in
the text is not shown.}
\label{measure}
\end{figure}

We will implement fault tolerant recovery using the method of Steane \cite
{Steane97,req}, which is based on preparing a $2n$ bit ancilla in a
superposition of $2^{n+k}$ product states which satisfy the parity checks in
the stabilizer. The measurement technique which underlies lemma 6 is
illustrated for a CSS code in figure \ref{measure}. In order to measure $%
\overline{X}_{010}$ in this example we prepare an ancilla in $\left| {000}
\right>_L + \left| {010} \right>_L$ and operate bitwise $^C\! X$, then
Hadamard transform the ancilla and measure it. This permits us to learn
simultaneously the result of measuring $\overline{X}_{010}$ on the logical
qubits, and the syndrome for $Z$ errors, which can then be corrected (the
whole network is repeated as necessary, see section \ref{s:5}). Replacing $%
^C\! X$ by $^C\! Z$, a measurement of $\overline{Z}_u$ can be accomplished
while learning simultaneously the syndrome for $X$ errors. The structure of
CSS codes permits the $2n$ bit ancilla to consist of two separate blocks of $%
n$ bits, which is why fig. \ref{measure} only shows an $n$ bit ancilla. For
general stabilizer codes the method is essentially the same but does not
have such an elegant expression in terms of logical qubit states.

A standard recovery, without measurement of any observable other than the
syndrome, involves the preparation of $\left| {0} \right>_L$. To prepare $%
\left| {0} \right>_L + \left| {u} \right>_L$ from $\left| {0} \right>_L$ we
simply add a one-bit Hadamard transform and $^C\! X$s to target bits at the
non-zero coordinates in $u \cdot D$. This is only slightly more complicated
than preparation of $\left| {0} \right>_L$ because the number of rows in $%
\tilde{H}$, which gives the network to construct $\left| {0} \right>_L$, is
much larger than 1 for powerful error correcting codes. Furthermore, since $%
\left| {0} \right>_L + \left| {u} \right>_L$ satisfies fewer parity checks
than $\left| {0} \right>_L$, the verification of the prepared state is
quicker. Hence the claim ``at no cost'' in lemma 6 is justified.

The final ingredient, before we can calculate error tolerance levels for
these methods, is to examine exactly how the switching/swapping and Toffoli
gates work, in order to see how frequently recovery must be performed.
Figures \ref{cnot} to \ref{switch} show example quantum networks. The
examples all show a case in which three logical qubits are stored in each
block, and horizontal lines indicate logical rather than physical qubits.
The zigzags on some of the qubits are a visual aid to keep track of the
quantum information, which propagates between blocks when operations such as
quantum teleportation take place. The figures show measurements of $%
\overline{X}$ or $\overline{Z}$ taking place by means of cat states. The
networks are drawn in this form to make it clearer how they operate, but it
is understood that at this point in the actual implementation the better
method of figure \ref{measure} and lemma 6 would be used, so a recovery
takes place. The exception is the $^{CC}\!\! Z$ gate in the network for the
Toffoli gate, figure \ref{Toff}, which must have a cat state as target. This
will be discussed shortly.

Figure \ref{cnot} shows a $^C\! X$ from the 2nd to the 3rd logical qubit in
a data block, using two ancilla blocks prepared in $\left| {000} \right>_L$.
The first part of the network is a quantum teleportation from the 2nd bit of
the data block to the 3rd bit of the 2nd ancilla. Then a block $^C\! X$
takes place from this ancilla onto the data block. Finally another
teleportation replaces the bit back into the data block. The whole operation
uses four recoveries.

Figure \ref{teleport} introduces a shorthand symbol for quantum
teleportation, and gives two example implementations, depending on which
ancilla states ($\left| {0}\right\rangle _{L}$ or $\left| {+}\right\rangle
_{L}$, etc.) one happens to have available. This is to make the point that
teleportation can be carried out via any state in the Bell basis, and so the
ancillas can be in one of many different initial states. This reduces the
amount of ancilla preparation needed for networks such as figures \ref{cnot}
and \ref{Toff}.

Figure \ref{Toff} gives the implementation of the Toffoli gate described by
Preskill \cite{Pres}, based on Shor's ideas. The figure does not show the
complete network. The operations in the dashed box are only carried out if
the measurement indicated gives a 1. If one or other of the two other
measurements on the data input block give a one, the network in the dashed
box changes, but it still involves simple whole-block operations plus two
teleportations. The other feature not shown on fig. \ref{Toff} is the
repetition of the measurement (via cat) used to prepare the ancillas. During
its preparation, the cat is verified against $X$ errors in order that the $%
^C\! \overline{Z}$ does not propagate uncorrectable errors into the 3rd
ancilla. However, the cat has a chance of acquiring a $Z$ error, which makes
the measurement fail with probability linear in the error rates. We
therefore repeat this part of the network, using further $n$-bit cats
prepared in parallel. By choosing a total of $d$ repetitions, where $d$ is
the minimum distance of the error-correcting code being used, we ensure that
by taking the majority vote, the probability of failure is lower than that
of accumulating too many errors during recovery. Overall, the Toffoli
network requires about 8 recoveries, allowing for two recoveries during the
ancilla preparation part of the network, one when the data block is
measured, two each for the teleportations, and a further one for the final
switching operation (see fig. \ref{switch}).

Figure \ref{switch} shows how to switch the $i$th logical bit between a data
block and an ancillary block, using a single recovery.

\section{Candidate quantum codes}

\label{s:4}

In this section we will find CSS codes which meet the requirements of the
fault tolerant methods considered in sections \ref{s:2} and \ref{s:3}.

The general idea is that we would like $C_{0}$ and its cosets given by $%
\tilde{D}$ to have weight distributions which permit methods such as lemma
1, while also forming the code $C_{0}^{\perp }$, in order to satisfy lemma
3. However, the possibilities are restricted by the fact that a self-dual
classical code over $GF(2)$ with weights all a multiple of $w>1$ can only
have $w=2$ or $w=4$ \cite{Turyn,MacWS}.

CSS codes $[[n=2^m-1,1,d]]$ satisfying lemma 1 with $w = 0 \mbox{ mod } 8$
can be constructed from punctured binary Reed-Muller codes, the simplest
example is $[[15,1,3]]$ given in \cite{KLZ96}. A related possibility is $%
[[n=2^m-1,k,d=3]]$, $m \ge 7$ in which $C_0$ is a punctured 1st order
Reed-Muller code (whose dual has minimum distance 3) and $C_0$ together with
its cosets make a punctured 2nd order Reed-Muller code. The simplest example
is $[[127,29,3]]$. The properties of these codes are far from optimal, so we
will not pursue them further.

Let us now concentrate on classical codes suitable for lemma 4 (and
therefore for all of lemmas 2--6). Let $C_0 = C^{\perp}$, then by the proof
of the lemma, all the weights of $C_0$ are multiples of 4. Such codes $C_0$
are called ``doubly even'', or ``type II'' or sometimes merely ``even.''
Doubly even codes are always contained in their duals since the rows of the
generator must all have even overlap with themselves and each other. We will
refer to the CSS codes having doubly even $C_0$ as ``lemma 4'' codes.

Note that once we have a $[[n,k,d]]$ lemma 4 code, a $[[n-1,k+1,d^{\prime}%
\ge d-1]]$ lemma 4 code can be obtained by deleting a row from the generator
of $C_0$ (which is the check matrix of $C$) \cite{Steane3}.

Any extended quadratic residue classical code of length $n=8m$ is doubly
even \cite{MacWS}. This yields a set of good lemma 4 codes beginning $%
[[24,0,8]]$, $[[48,0,12]]$, $[[80,0,16]]$, $[[104,0,20]]$ \ldots , in this
list only the code of the smallest $n$ for given $d$ is mentioned. No better
codes exist for $d<20$, and none better is known to exist for $d=20$ \cite
{selfdual}. From these we obtain lemma 4 codes such as $[[47,1,11]]$, $%
[[79,1,15]]$, $[[99,5,15]]$ which are considered in section \ref{s:5}. In
section \ref{s:5} (eq. (\ref{s})) we will need to know enough about the
weight distribution of $C_{0}$ to estimate the average weight of a row of
the generator matrix of $C_{0}$. This average weight is equal to the average
weight of the generator of the self-dual code we started from, which is well
approximated by its minimum distance in this case (i.e. respectively
8,12,16,20 for the codes in the first list in this paragraph).

Bose Chaudhuri Hocquenghem (BCH) classical codes \cite{BC,H,MacWS} yield a
good set of CSS quantum codes. The condition for a BCH code to contain its
dual was discussed in \cite{Grassl,eCSS}. It can be shown that the dual of
double- and triple-error correcting BCH codes of length $2^m-1$ is doubly
even when $m>3$ \cite{MacWS}, and I conjecture that the dual of a BCH code
of length $2^m-1$ is doubly even whenever the code contains its dual. I have
checked that this conjecture is satisfied for $n \le 127$ by examining the
parity check matrices. Hence we have a large class of lemma 4 codes,
containing for example $[[31,11,5]]$, $[[31,1,7]]$, $[[63,39,5]]$, $%
[[63,27,7]]$, $[[127,85,7]]$, $[[127,43,13]]$, $[[127,29,15]]$, and $%
[[255,143,15]]$ by conjecture. Of these, the codes $[[127,29,15]]$ and $%
[[127,43,13]]$ yield the best results in section \ref{s:5}. For length $2^m-1
$ BCH codes, the weights of the rows of the parity check matrix are $2^{m-1}$%
. This value is required in section \ref{s:5} (equation (\ref{s})).

\section{Error tolerance and overheads}

\label{s:5}

We now wish to estimate the amount of noise which can be tolerated by a
quantum computer using the methods discussed. The estimate is made through
an analysis based on that in \cite{req}, but with some new features.

\begin{figure}[tbp]
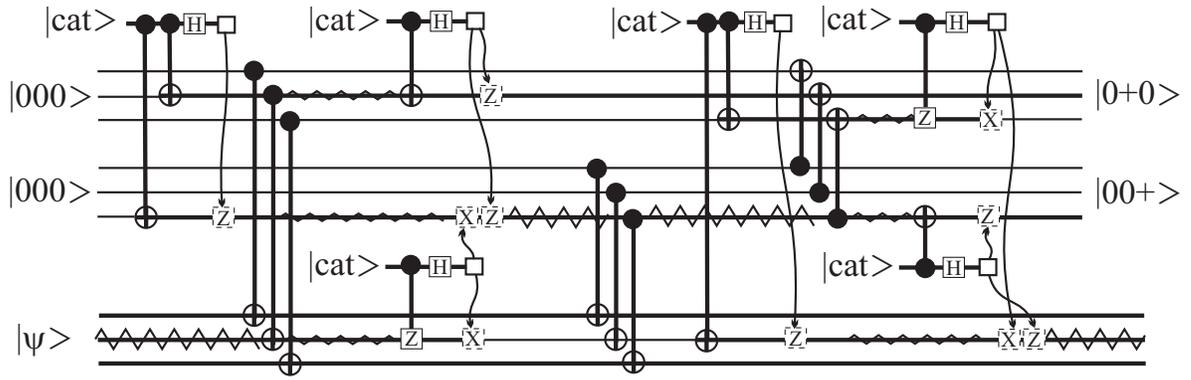

\caption{ Network to perform $^C\! \overline{X}$ between two logical qubits
in the same block, in this case the 2nd and 3rd qubits of the lower block.
The horizontal lines indicate logical qubits, in this example there are
three logical qubits per block. A small empty box represents a measurement.
An operator in a dashed box indicated by an arrow is only carried out if the
relevant measurement yields a 1. The zigzag lines are a visual aid to help
keep track of the quantum information which moves between blocks when
quantum teleportations take place. The horizontal lines are shown narrow
when the relevant qubit is in the state $\left| {0} \right>_L$.}
\label{cnot}
\end{figure}

\begin{figure}[tbp]
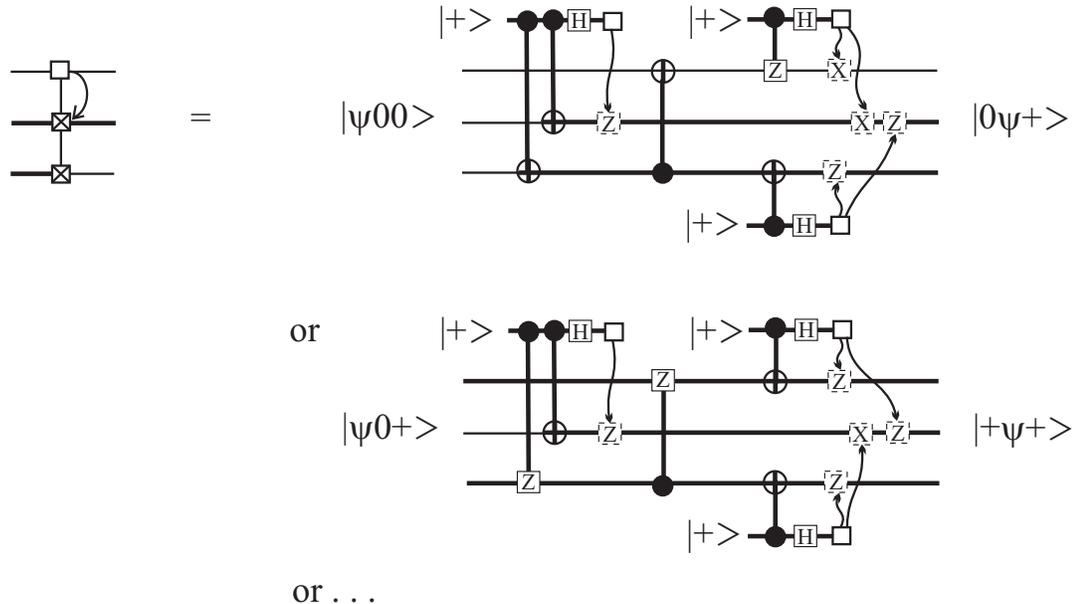

\caption{ Illustration of two ways to implement teleportation. The symbol on
the left is a shorthand which is used in fig. 4. The two crossed boxes
indicate the qubits which are placed in a Bell state, and the arrow
indicates whence and whither the qubit is teleported.}
\label{teleport}
\end{figure}

\begin{figure}[tbp]
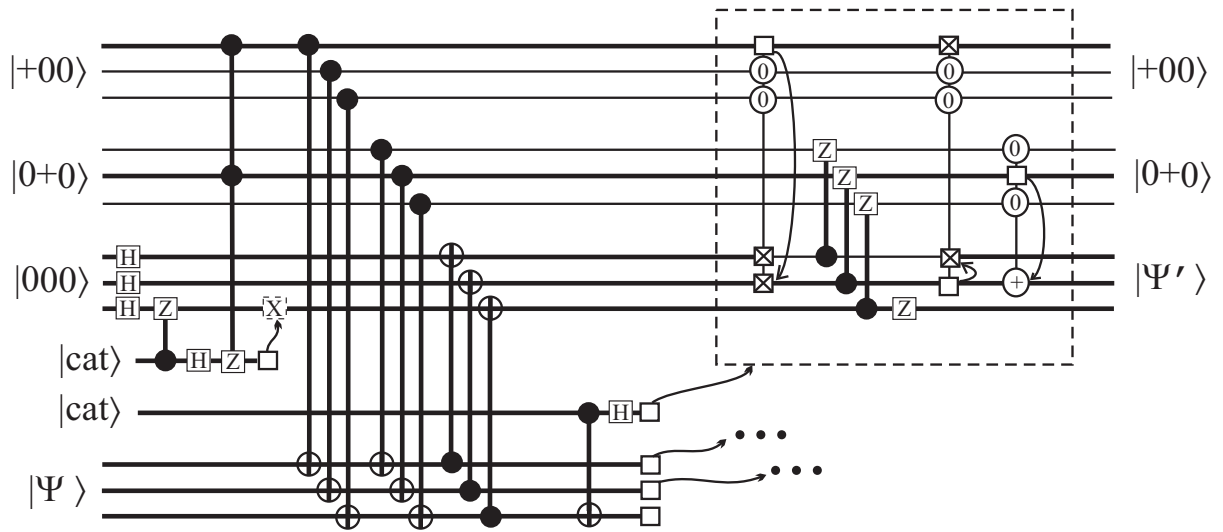

\caption{Fault-tolerant Toffoli gate on three qubits in the same block. If
the data block at the bottom is initially in the state $\left| {\Psi}
\right>_L$, then the third ancilliary block ends up in the state $\overline{T%
} \left| {\Psi} \right>_L$. The operations in the dashed box are only
carried out if the indicated measurement yields a 1. The figure does not
show further operations which are required if the other measurements yield
1s, nor the repetition of the initial ``measurement via cat'' which prepares
the ancillas (see text). The encircled zeros and the encircled plus sign are
a visual reminder that those qubits are in the states $\left| {0} \right>_L$
and $\left| {+} \right>_L$, respectively, which allows the teleportation and
switching operations to function.}
\label{Toff}
\end{figure}

\begin{figure}[tbp]
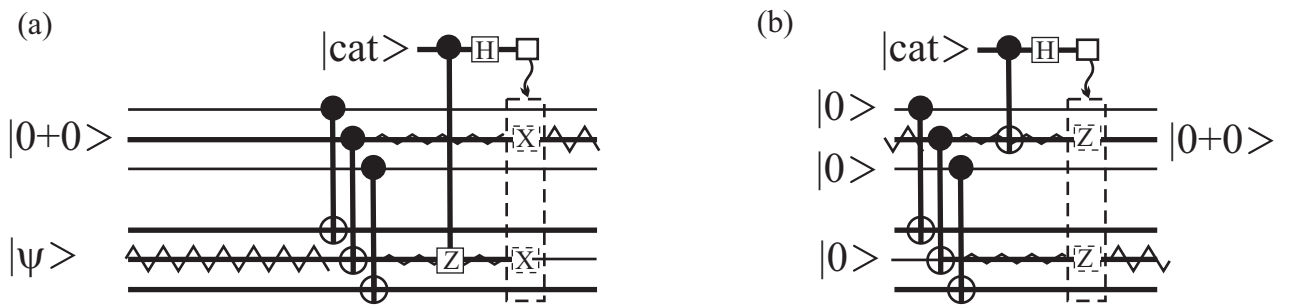

\caption{Switching the $i$'th bit out of a block (a), and back in again (b).}
\label{switch}
\end{figure}

The quantum computer will operate as follows. A computation involving $K$
logical qubits will be carried out using $K/k$ blocks to store the quantum
information, plus a further $3$ blocks which act as an ``accumulator''. Each
accumulator block uses the same error correcting code as the rest of the
computer, but only stores one logical qubit at a time. Quantum gates on the
logical information in the computer are carried out via the accumulator
using the methods illustrated in figs \ref{cnot} to \ref{switch}. A larger
accumulator of $\sim K/4k$ blocks could be used without greatly changing the
overall results.

In order to extract syndromes, we prepare, in parallel, 4 ancillary blocks
for every data block or accumulator block in the computer. A single complete
syndrome is extracted for the $3+K/k$ blocks, using 2 of the prepared
ancillas for each block. Whenever this first syndrome indicates no errors,
we accept it even though it has a small chance of being the wrong syndrome 
\cite{Zalka}. Those blocks are left alone, any errors they contain will with
high probability be corrected by the next recovery. There remain a number of
blocks whose first syndrome was non-zero. Each of these blocks will be
corrected, but only after the syndrome has been extracted a further $O(d)$
times, and the best estimate syndrome (e.g. by majority vote) is used to
correct the block. The ancillas required for these further extractions have
already been prepared: they are the remaining $2(3+K/k)$ ancillas which were
not used for the first syndrome.

Each ancilla uses $n$ qubits to store the prepared state, plus one used for
verification, therefore the total number of physical qubits in the computer
is $(5n+4)(3 + K/k)$. The ratio 
\begin{equation}
S = \frac{5n+4}{k} \left(1 + 3k/K \right)  \label{scaleup}
\end{equation}
is the scale-up in computer size necessary to allow fault tolerance by this
method.

The method described in the previous paragraphs makes better use of ancillas
than the simplest approach of generating $O(d)$ syndromes for every block.
In order that the prepared ancillas are sufficient in number, we require
that the probability of obtaining a non-zero syndrome is less than $O(1/d)$.
We will confirm that this is the case at the end of the calculation.

We need to estimate the number of elementary gates and time steps required
to prepare an ancillary $n$-bit block in the state $\left| {0}\right\rangle
_{L}$, and to verify the state so that the only $X$ errors which remain in
it are uncorrelated with each other. Let $\tilde{H}$ be the generator of $%
C_{0}$, and let $w$ be the average weight of a row of $\tilde{H}$. We would
like $w$ to be small so that ancillas can be prepared quickly but the
construction of $[[n,k,d]]$ CSS codes rules this out for $k\gg 1$. Since all
the cosets used to build the quantum codewords (see eq. \ref{uL}) are at
distance at least $d$ from each other, $C_{0}$ must consist of words
separated by significantly more than $d$, so $w$ may be several times larger
than $d$.

The network to build $\left| {0}\right\rangle _{L}$ consists of one Hadamard
gate and $w-1$ controlled-nots for each row of $\tilde{H}$. In \cite{req} it
was assumed that gates on different blocks could be performed in parallel,
while 2- or more-bit gates within a block could not. Here we shall allow a
further type of parallelism, namely that a multiple controlled-not,
involving one control bit and several target bits in the same block, can be
performed in a single time step. This is physically reasonable since it is
possible in implementations such as the ion trap, in which a communal degree
of freedom is coupled to every qubit in a block, and we can drive
transitions in many qubits simultaneously. The number of time steps required
to build $\left| {0}\right\rangle _{L}$ is therefore equal to the number of
rows in $\tilde{H}$, which is $(n-k)/2$.

The most time-consuming part of the ancilla preparation is the verification.
This involves the evaluation of parity checks using controlled nots from
several control bits to a single target bit, which cannot be done in
parallel. We will assume a thorough verification, evaluating all the parity
checks in $\tilde{H}$ and $\tilde{D}$. The former confirms that the prepared
state is in the encoded Hilbert space, the latter that $\left| {0} \right>_L$
rather than some other encoded state has been prepared. To keep the number
of time steps to a minimum, we arrange for $\tilde{D}$ to contain as few 1s
as possible. Since we have a distance $d$ CSS code, the weight of each row
of $\tilde{D}$ is at least $d$. We will assume $\tilde{D}$ can be arranged
to have a mean weight per row of $d+1$. Hence the total number of $^C\! X$
gates used for verification of one ancilla is $w(n-k)/2 + (d+1)k$.

One complete recovery of the computer fails if any block develops more than $%
t \equiv (d-1)/2 $ errors, or any correction is applied on the basis of a
wrong syndrome. It is shown in \cite{req} that the failure probability is
dominated by the former when the syndrome is extracted $O(t+1)$ or more
times. First consider the blocks whose first extracted syndrome was
non-zero. For one of these blocks the probability to develop more than $t$
errors is \cite{req} 
\begin{equation}
P \simeq 2 \sum_{i=t+1}^g \frac{g!}{i!(g-i)!} \left( \frac{2}{3}\gamma + 
\frac{s}{g} \frac{2}{3}\epsilon \right)^i  \label{P}
\end{equation}
where $\gamma$ is the probability of gate failure, $\epsilon$ the
probability of memory error per time step, and $g$,$s$ are the number of
independent opportunities for gate, memory errors respectively. The errors
either occur directly in the block to be corrected, or they originate in an
ancilla and are subsequently propagated into the block. Using the
assumptions made above concerning the weights of $\tilde{H}$ and $\tilde{D}$%
, and the degree of parallelism, an analysis similar to that in \cite{req}
yields 
\begin{eqnarray}
g &\simeq& n(4r+1)  \label{g} \\
s &\simeq& n\left( \left( w+2 \right)\left( \frac{n-k}{2} \right) + (d+2)k +
n(2+r/2) \right),  \label{s}
\end{eqnarray}
where $r$ is the number of repetitions of the syndrome extraction required
on average for confidence that one has the right syndrome. We will take $r =
t+1$, which is a safe over-estimate, as discussed in \cite{req}.

The only possibility left out of equation (\ref{P}) is that a zero syndrome
is wrongly obtained for some block during the first round of syndrome
extraction (thus the block remains uncorrected), and then the further errors
developed during the next recovery bring the total up to more than $t$
before correction is applied. Consider the two successive recoveries of such
a block, in which no corrective measure is applied after the first recovery.
There are approximately twice as many error opportunities, so the
probability that $t+1$ errors develop can be estimated as $2^{t+1} P$.
However, to wrongly obtain a zero syndrome requires an error in the syndrome
which matches the syndrome, which is highly unlikely. Its probability is
small compared to $1/2^{t+1}$, therefore this failure mechanism is much less
likely than the one leading to eq. (\ref{P}).

To run a quantum algorithm needing $K$ qubits and $Q$ Toffoli gates, we
require 
\begin{equation}
P < \frac{k}{8 K Q}  \label{Plim}
\end{equation}
since we require $8 Q$ recoveries of each block (see section \ref{s:3}) and
there are $3+K/k \simeq K/k$ blocks, and all these recoveries must succeed
so that the overall success probability is greater than a half.

Table 1 shows the scale-up and error rates needed to satisfy (\ref{P}) to (%
\ref{Plim}) for various quantum error correcting codes, for $K Q = 2.15
\times 10^{12}$. This size of computation is sufficient to factorize a 130
digit (430 bit) number using Shor's algorithm \cite{Beck,Pres}. Comparing
the $[[127,29,15]]$ and $[[127,43,13]]$ codes with the $[[47,1,11]]$ and $%
[[79,1,15]]$ codes we see that the more efficient codes allow one to save
about a factor 10 in the scale-up $S$, with no change in noise level, or a
factor 17 in $S$ if the memory noise $\epsilon$ is reduced by a factor 3. If
we wish to factorize thousand-digit numbers, then $K Q$ grows by a factor $%
9^4 = 3^8$. However, $P$ scales as $\gamma^{t+1}$ so (\ref{Plim}) would
still be satisfied by the distance 15 codes if $\gamma$ and $\epsilon$ were
reduced by a factor 3.

The assumption that eq. (\ref{scaleup}) allows for sufficiently many
ancillas needs to be verified. The probability $P_1$ of obtaining a non-zero
syndrome can be estimated using eq. (\ref{P}) but letting the sum run from 1
to $g$ instead of $t+1$ to $g$. This gives $P_1 \simeq 32 n r \gamma / 3$
when $\epsilon = \gamma/n$. Examining table 1, we find $P_1$ is largest for
the distance 15 codes having $n=127$ and $255$, which give respectively $P =
0.25, 0.31$ using $\gamma=2 \times 10^{-5}$, $1.1 \times 10^{-5}$
respectively. The former case is satisfactory, since if there are $B$ blocks
in the computer, then after the first syndrome extraction we have $2B$
available ancillas and we only need $7 \times 0.25 B$ to complete the 8-fold
repetition of extraction of the non-zero syndromes. The latter case ($P_1 =
0.31$) is not satisfactory, but becomes so if we reduce $\gamma$ to $1
\times 10^{-5}$. This confirms eq. (\ref{scaleup}).

To conclude, fault tolerant quantum computing can work well with efficient
quantum error correcting codes such as the $[[127,29,15]]$ CSS code obtained
from the classical $[127,78,15]$ BCH code. The success relies on the ability
to merge useful measurements on the logical state with recovery operations,
on careful network design and on optimized use of ancillas. These insights
allow large quantum computations, such as factorization of 100 or 1000-digit
numbers, to proceed on a quantum computer about an order of magnitude
smaller than previously thought, without change in the necessary noise
level. The fault tolerant quantum computer need only be about one order of
magnitude larger than the logical computer contained within it.

The author is supported by the Royal Society and by St. Edmund Hall, Oxford.

\newpage

\begin{tabular}{ccccc}
code & $P$ & $\gamma$ & $\epsilon$ & $(5n+4)/k$ \\ 
& $(10^{-14})$ & $(10^{-6})$ & $(10^{-6})$ &  \\ \hline
$[[99,5,15]]$ & 29 & 28 & 0.28 & 100 \\ 
$[[127,29,15]]$ & 169 & 20 & 0.16 & 22 \\ 
$[[255,143,15]]$ & 831 & 11 & 0.04 & 9 \\ 
&  &  &  &  \\ 
$[[127,43,13]]$ & 250 & 13 & 0.10 & 15 \\ 
&  &  &  &  \\ 
$[[63,27,7]]$ & 157 & 1.4 & 0.02 & 12 \\ 
&  &  &  &  \\ 
$[[47,1,11]]$ & 5.8 & 14 & 0.30 & 239 \\ 
$[[79,1,15]]$ & 5.8 & 30 & 0.38 & 399
\end{tabular}

Table 1. Error rates and scale up required to run a quantum algorithm of
size $K Q = 2.15 \times 10^{12}$. The first column gives the parameters of
the quantum code, which is identified in section \ref{s:4}. The 2nd column
gives the required success probability for recovery of a single block (eq. (%
\ref{Plim})). The 3rd and 4th columns give the required error rates (eqs. (%
\ref{P})--(\ref{s})), and the final column gives the scale-up in computer
size (eq. (\ref{scaleup})).

\newpage

\end{document}